\documentstyle{article}
\title{Dimensional reduction of
quantum fields on a brane}
\author{ Z. Haba\\Institute of Theoretical Physics, University of Wroclaw,
\\50-204 Wroclaw, Plac Maxa Borna 9, Poland\\e-mail:zhab@ift.uni.wroc.pl}
\date{}
\begin{document}
\maketitle
\begin{abstract}
If we restrict a quantum field defined on a regular $D$
dimensional curved manifold to a  $d$ dimensional submanifold then
the resulting field will still have the singularity of the
original $ D $ dimensional model. We show that a singular
background metric can force the restricted field to behave as a
$d$ dimensional  quantum field.
\end{abstract}
\section{ Introduction}
Quantum fields are defined by their correlation functions. The
Lagrangian serves as a heuristic tool for a construction of
quantum fields. A reduction of the number of  coordinates in the
$D$ dimensional Lagrangian  does not mean that if we had a
complete $D$-dimensional quantum field theory then we could reduce
it in any way to a model resembling a quantum field theory in
$d<D$ dimensions. We can see this problem already at the level of
a massless free field $\phi$. The vacuum correlation function of
$\phi({\bf x}(1)) $ and $\phi({\bf x}(2))$ is $\vert {\bf
x}(1)-{\bf x}(2)\vert^{-D+2}$. If we restrict the field to the
hypersurface $x_{D}=0$ setting in all correlation functions
$x_{D}(j)=0$ then we obtain a quantum  field with a continuous
mass spectrum in $d=D-1$ dimensions but this will not be the
canonical free field in $d$ dimensions whose two-point function
behaves as $\vert {\bf x}(1)-{\bf x}(2)\vert^{-D+3}$ at short
distances. Nevertheless, it is an attractive idea that the
Universe once had more dimensions and subsequently through a
dynamical process shrank to a lower dimensional hypersurface. The
dynamics could have the form of a gravitational collapse ( say a
ball collapsing to a disk). At the level of field correlation
functions this would mean that we have initially scalar,
electromagnetic and gravitational fields in $D$-dimensions with
their standard canonical singularities which subsequently evolve
into fields with
 $d<D$ dimensional singularity. We show
 that such a reduction of  dimensions is possible
 when the metric becomes singular.
A similar mechanism is suggested in the brane scheme of
refs.\cite{dvali}\cite{rs}. In ref.\cite{dvali} the authors derive
the Green's function in $D=5$ dimensional space-time which on the
$d=D-1=4$ submanifold has the singularity of the fourdimensional
Green's function. Their model encounters some difficulties when
generalized to arbitrary $D$ and $d$ \cite{dvali2}. Some other
brane-type models of quantum fields are discussed in
refs.\cite{rubakov} \cite{dvali3}\cite{mintchev}\cite{giddings}.
In this letter we discuss a general metric which has power-law
singularity. In general relativity such metrics could  describe
collapse phenomena \cite{szekeres}. We can obtain  metrics with
power-law singularities as solutions to higher dimensional
supergravity theories \cite{super}. These solutions describe $m$-
branes or intersecting m-branes in an $m+n $ dimensional
space-time \cite{sing}).
\section{A quantum field on a D-1 dimensional hypersurface}
We consider a submanifold ${\cal M}_{D-1}$ of a Riemannian
manifold ${\cal M}_{D}$ whose metric becomes singular near ${\cal
M}_{D-1}$. The metric on ${\cal M}_{D}$  close to ${\cal M}_{D-1}$
(in local coordinates) is described by a "warp factor" $ a(x_{D})$
which becomes singular either when $x_{D}\rightarrow 0$ or $\vert
x_{D}\vert\rightarrow \infty$
\begin{equation}
 ds^{2}\equiv g_{\mu\nu}dx^{\mu}dx^{\nu}=dx_{D}^{2}+a( x_{D})^{2} (dx_{1}^{2}
 +....+dx_{D-1}^{2})
 \end{equation}
 The Green's function of the minimally coupled scalar field is
 a solution of the equation
\begin{equation}
{\cal A}G=g^{-\frac{1}{2}}\delta
\end{equation}
where ${\cal A}$ is the Laplace-Beltrami operator
\begin{equation}
{\cal
A}=g^{-\frac{1}{2}}\partial_{\mu}(g^{\mu\nu}g^{\frac{1}{2}}\partial_{\nu})
\end{equation}
In the metric (1) eq.(2) reads
\begin{equation}
(\partial_{D}a^{D-1}\partial_{D}
+a^{D-3}\triangle)G=\delta(x_{D}-x_{D}^{\prime}) \delta({\bf
x}-{\bf x}^{\prime})
\end{equation}
where $d=D-1$ , ${\bf x}=(x_{1},....,x_{d})$ and $\triangle$ is
the $d$-dimensional Laplacian. This equation is simplified if we
introduce the coordinate
\begin{equation}
\eta=\int a^{-d}dx_{D}
\end{equation}
Then
\begin{equation}
(\partial_{\eta}^{2}+a^{2d-2}\triangle)G=\delta(\eta-\eta^{\prime})
\delta({\bf x}-{\bf x}^{\prime})
\end{equation}
In the paper of Dvali et al \cite{dvali}$D=4$ and
$a^{4}(x_{D}(\eta))\rightarrow\delta(\eta)$.

 We discuss in detail the case
\begin{equation}
 a(x_{D})=\vert x_{D}\vert^{\alpha}
 \end{equation}
 Then
\begin{equation}
(\partial_{D}\vert x_{D}\vert^{\alpha d}\partial_{D} +\vert
x_{D}\vert^{\alpha(d-2)}\triangle)G=\delta(x_{D}-x_{D}^{\prime})
\delta({\bf x}-{\bf x}^{\prime})
\end{equation}
We define
\begin{equation}
\eta=\vert 1-\alpha d\vert^{-1}x_{D}\vert x_{D}\vert^{-\alpha d}
\end{equation}
then eq.(8) takes the form
\begin{equation}
(\partial_{\eta}^{2}+\kappa\vert\eta\vert^{2\nu}\triangle)G=\delta(\eta-\eta^{\prime})
\delta({\bf x}-{\bf x}^{\prime})
\end{equation}
or in terms of the Fourier transform $\tilde{G}$ in ${\bf x}$
\begin{equation}
(\partial_{\eta}^{2}-{\bf
p}^{2}V(\eta))\tilde{G}=\delta(\eta-\eta^{\prime})
\end{equation}
where \begin{equation}
V(\eta)=\kappa\vert\eta\vert^{2\nu}\end{equation}
with
\begin{equation} \nu=\alpha (d-1)(1-\alpha d)^{-1}
\end{equation}
and \begin{displaymath}\kappa=\vert 1-\alpha d\vert^{-2\nu}
\end{displaymath}
 Eq.(11) can be solved by means of the Feynman-Kac integral
applying the proper time method
\begin{equation}\begin{array}{l} G(\eta,{\bf x};\eta^{\prime},{\bf
x}^{\prime})=\frac{1}{2}(2\pi)^{-d} \int_{0}^{\infty}d\tau\int
d{\bf p}
         \exp\left(i{\bf p}\left({\bf x}^{\prime}-{\bf
x}\right)\right)
 \cr E[\delta\left(
\eta^{\prime}-\eta- b\left(\tau\right)\right)
\exp\left(-\frac{1}{2}{\bf p}^{2}\int_{0}^{\tau}
V\left(\eta+b\left(s\right)\right)ds\right)]\end{array}
\end{equation}
Here, $b(s)$ is the Brownian motion \cite{simon} defined as the
Gaussian process with the covariance
\begin{equation}
E[b(s)b(t)]=min(s,t)
\end{equation}
$E[..]$ denotes an average over the paths of the Brownian motion.

 The dimensional reduction is imposed
by setting $\eta=\eta^{\prime}=0$. Next, we use the equivalence
$b(s)=\sqrt{\tau}b(\frac{s}{\tau})$ which follows from eq.(15).
Then, using the scaling invariance of the potential V
(i.e.,$V(\lambda\eta)=\lambda^{2\nu}V(\eta)$) we have
 \begin{equation}
\begin{array}{l}
G(0,{\bf x};0,{\bf x}^{\prime})=\frac{1}{2}(2\pi)^{-d}
\int_{0}^{\infty}d\tau\int d{\bf p}
         \exp\left(i{\bf p}\left({\bf x}^{\prime}-{\bf
x}\right)\right)
 \cr E[\delta\left(
\sqrt{\tau}b\left(1\right)\right) \exp\left(-
\frac{1}{2}\tau^{1+\nu}{\bf p}^{2}\int_{0}^{1}
V\left(b\left(s\right)\right)ds\right)]\end{array}
\end{equation}
Changing the variables
\begin{displaymath}
{\bf p}=\tau^{-\frac{1}{2}-\frac{\nu}{2}}{\bf k}
\end{displaymath}
and
\begin{displaymath}
\tau=r\vert {\bf x}-{\bf x}^{\prime}\vert^{\frac{2}{1+\nu}}
\end{displaymath}
we obtain
\begin{equation}
G(0,{\bf x};0,{\bf x}^{\prime})=C\vert{\bf x}^{\prime}-{\bf
x}\vert^{-d+\frac{1}{1+\nu}}
\end{equation}
with a certain constant $C$. If $0>\nu>-1$ then the singularity of
the Green's function is weaker  than the one for the
$D$-dimensional free field. The Green's function is equal to the
Green's function of the $ d=D-1$ dimensional free field if
$\nu=-\frac{1}{2}$ what corresponds to $\alpha=\frac{1}{2-d}$. The
potential with $2\nu=-1$ has the same scaling dimension as
$V=\delta(\eta)$ applied by Dvali et al \cite{dvali}. The
Hamiltonian with the potential $V(\eta)=\vert\eta\vert^{-1}$ and
the path integral (16) require a careful definition if $2\nu\leq
-1$ but at least till $2\nu \geq -2$ such a definition (through a
regularization and a subsequent limiting procedure) is possible
\cite{klauder}. Eq.(11) with the $\delta$-potential
("$\delta$-brane") also involves a particular regularization and
its subsequent removal \cite{albeverio}.
 Let us
consider a solution of this problem by means of the proper time
method. The heat kernel $K^{\delta}$ is known exactly for the
$\delta$-potential \cite{gaveau}. Hence,
\begin{equation}
\begin{array}{l} G(\eta,{\bf x};\eta^{\prime},{\bf
x}^{\prime})
 =\frac{1}{2}(2\pi)^{-4} \int_{0}^{\infty}d\tau\int d{\bf p}
         \exp\left(i{\bf p}\left({\bf x}^{\prime}-{\bf
x}\right)\right)K^{\delta}(\eta,\eta^{\prime},\tau)\cr
=\frac{1}{2}(2\pi)^{-4} \int_{0}^{\infty}d\tau\int d{\bf p}
         \exp\left(i{\bf p}\left({\bf x}^{\prime}-{\bf
x}\right)\right)
 \cr \Big(
K_{0}(\eta-\eta^{\prime},\tau)-2{\bf
p}^{2}\int_{0}^{\infty}du\exp(-2{\bf p}^{2}u)K_{0}(\vert \eta\vert
+\vert \eta^{\prime}\vert+u,\tau)\Big)
 \end{array}
\end{equation}
where \begin{equation} K_{0}(\eta,\tau)=(2\pi\tau)^{-\frac{1}{2}}
\exp(-\frac{1}{2\tau}\eta^{2})
\end{equation}
is the heat kernel for the Brownian motion.

 When
$\eta=\eta^{\prime}=0$ the $\tau$-integral of the first term on
the r.h.s. of eq.(18) (the one independent of ${\bf p}$ ) is
infinite (and proportional to $\delta({\bf x}-{\bf x}^{\prime})$)
whereas the second integral gives the formula(17) with
$\nu=-\frac{1}{2}$.

Eq.(18) could have been derived as a limiting case of eqs. (6)
and(16) when $a(x_{D})^{2d-2}\rightarrow\delta(\eta)$. On the
Lagrangian level we have
\begin{equation}
\int dx_{D}d{\bf x}\sqrt{g}g^{DD}\partial_{D}\phi\partial_{D}\phi
=\int d\eta d{\bf x}\partial_{\eta}\phi\partial_{\eta}\phi
\end{equation}
and
\begin{equation}
\begin{array}{l} \int dx_{D}d{\bf
x}\sqrt{g}g^{jk}\partial_{j}\phi\partial_{k}\phi =\int d\eta d{\bf
x}a^{2d-2}\partial_{j}\phi\partial_{j}\phi  \rightarrow \int d\eta
d{\bf x}\delta(\eta)\partial_{j}\phi\partial_{j}\phi
\end{array}
\end{equation}
Hence, we recover the Lagrangian of Dvali et al $\cite{dvali}$.
\section{A generalization to  surfaces of arbitrary dimensions}
Let us consider on a $D=m+n$ dimensional manifold a metric (in
local coordinates)  which close to the $n$-dimensional surface
takes the form
\begin{equation}
ds^{2}=\vert{\bf y}\vert^{2\beta}d{\bf y}^{2}+\vert {\bf
y}\vert^{2\alpha}d{\bf x}^{2}
\end{equation}
where ${\bf y}\in R^{m}$ and ${\bf x}\in R^{n}$. Eq.(2) for the
Green's function of the Laplace-Beltrami operator reads
\begin{equation} \Big(\frac{\partial}{\partial y^{i}}\vert{\bf
y}\vert^{\beta(m-2)+\alpha n}\frac{\partial}{\partial y^{i}}+\vert
{\bf y}\vert^{\beta m+\alpha(n-2)}\frac{\partial}{\partial
x^{j}}\frac{\partial}{\partial x^{j}}\Big)G_{E}=\delta
\end{equation}
We discuss here only a simplified form of eq.(23) which appears
when\begin{equation} \beta(m-2)+\alpha n=0
\end{equation}
In such a case eq.(23) reads
\begin{equation}
\Big(\frac{\partial}{\partial y^{i}}\frac{\partial}{\partial
y^{i}}+\vert {\bf y}\vert^{2\beta
-2\alpha}\frac{\partial}{\partial x^{j}}\frac{\partial}{\partial
x^{j}}\Big)G_{E}=\delta\end{equation}  or taking the Fourier
transform in ${\bf x}$
\begin{equation} \Big(\frac{\partial}{\partial
y_{i}}\frac{\partial}{\partial y_{i}}-{\bf p}^{2}V( {\bf
y})\Big)\tilde{G}_{E}=\delta({\bf y})
\end{equation}
We obtain again  an equation for the Green's function of the
Schr\"odinger operator  with the potential
\begin{equation}
V({\bf y})=\vert {\bf y}\vert^{2\beta-2\alpha}
\end{equation}
and  the coupling constant ${\bf p}^{2}$. We solve eq.(26) by
means of the proper time method \begin{equation}\begin{array}{l}
G({\bf y},{\bf x};{\bf y}^{\prime},{\bf
x}^{\prime})=\frac{1}{2}(2\pi)^{-n} \int_{0}^{\infty}d\tau\int
d{\bf p}
         \exp\left(i{\bf p}\left({\bf x}^{\prime}-{\bf
x}\right)\right)
 \cr E[\delta\left(
{\bf y}^{\prime}-{\bf y}- {\bf b}\left(\tau\right)\right)
\exp\left(-\frac{1}{2}{\bf p}^{2}\int_{0}^{\tau} V\left({\bf
y}+{\bf b}\left(s\right)\right)ds\right)]\end{array}
\end{equation}
where ${\bf b}$ is the $m$-dimensional Brownian motion.

 On the
brane ${\bf y}={\bf y}^{\prime}={\bf 0}$. In such a case using
${\bf b}(s)=\sqrt{\tau}{\bf b}(\frac{s}{\tau}) $ we have
\begin{equation} \int_{0}^{\tau}ds V({\bf
b}(s))=\tau^{1+\beta-\alpha}\int_{0}^{1} V({\bf b}(s))ds
\end{equation} Hence, if we change variables
\begin{displaymath}
{\bf p}={\bf k}\tau^{-\frac{1}{2}(1+\beta-\alpha)}
\end{displaymath}
then
\begin{equation}\begin{array}{l}
G({\bf 0},{\bf x};{\bf 0},{\bf x}^{\prime})G({\bf 0},{\bf x};{\bf
0},{\bf x}^{\prime}) \cr =\frac{1}{2} (2\pi)^{-n}
\int_{0}^{\infty}d\tau \sqrt{\tau}^{n(\alpha-1-\beta)-m}\int d{\bf
k}
         \exp\left(i{\bf k}\sqrt{\tau}^{\alpha-1-\beta}\left({\bf x}^{\prime}-{\bf
x}\right)\right)
 \cr E[\delta\left(
{\bf b}\left(1\right)\right) \exp\left(-\frac{1}{2}{\bf
k}^{2}\int_{0}^{1} V\left({\bf b}\left(s\right)\right)ds\right)]=C
\vert {\bf x}-{\bf x}^{\prime}\vert^{-n+\rho}\end{array}
\end{equation}with a certain constant $C$
and
\begin{displaymath}
\rho=(2-m)(1-\alpha+\beta)^{-1}
\end{displaymath}
For canonical quantum fields in $n$ dimensions we should have
$\rho=2$. This happens if (in addition to eq.(24))\begin{equation}
\alpha-\beta=\frac{m}{2}
\end{equation}
In such a case the potential is
\begin{equation}
V({\bf y})=\vert {\bf y}\vert^{-m}
\end{equation}
The potential (32) scales in the same way as the $\delta$-function
in $m$-dimensions. This is a singular potential. However, its
regularization $V_{\epsilon}({\bf y})=\vert {\bf
y}\vert^{-m-\epsilon}$ for any $\epsilon>0$ gives a self-adjoint
Hamiltonian  with the well-defined path integral. As $\epsilon$
can be arbitrarily small the Newton potential on the brane would
be indistinguishable from $r^{-1}$ if the brane is $n-1=3$
dimensional. We could again consider the limit $V({\bf
y})\rightarrow \delta({\bf y})$ in order to derive the model of
Dvali et al \cite{dvali2}. In contradistinction to the case $m=1$
the models in $m>1$ dimensions are more complicated. For $m=2$ and
$m=3$ the relation of the coupling constant ${\bf p}^{2}$ in
eq.(26) to the parameters appearing in the heat kernel
$K^{\delta}$ is not so explicit \cite{brzezniak}. For $m>3$ the
$\delta$-potential cannot be defined at all \cite{faddeev}
\cite{albeverio}.

We have discussed only scale invariant metrics . If the metric is
not scale invariant but its asymptotic behaviour for ${\bf
y}\rightarrow 0$ is of the form (22) then our results hold true
when $-1\leq \nu\leq 0$ and when applied to the short distance
behaviour $\vert {\bf x}-{\bf x}^{\prime}\vert\rightarrow 0$ of
$G({\bf 0},{\bf x};{\bf 0},{\bf x}^{\prime})$. If the asymptotic
behaviour of the metric for $\vert {\bf y}\vert \rightarrow \infty
$ is of the form (22) then our results apply if $\beta\geq\alpha$
to the behaviour of the Green's functions $G({\bf 0},{\bf x};{\bf
0},{\bf x}^{\prime})$  for large $\vert {\bf x}- {\bf
x}^{\prime}\vert $. In such a case $\rho<2$ in eq.(30), hence
$G({\bf 0},{\bf x};{\bf 0},{\bf x}^{\prime})$ and the
gravitational potential decay to zero faster than in the Newton
theory (in the $n$ dimensions). Depending on the asymptotic
behaviour of the metric tensor $g({\bf x},{\bf y})$ we obtain
models which lead to a modification of the Newton law either at
small or at large distances ( some brane models modifying the
classical gravity at small or large distances are discussed in
\cite{rubakov}\cite{giddings}).

\end{document}